\documentclass[9pt,twocolumn,twoside]{opticajnl}

\journal{opticajournal} 

\setboolean{shortarticle}{true}

\usepackage{lineno}

\title{Virtual Atom-Photon Bound States and Spontaneous Emission Control}

\author[1,2,*]{Stefano Longhi}

\affil[1]{Dipartimento di Fisica, Politecnico di Milano, Piazza L. da Vinci 32, I-20133 Milano, Italy}
\affil[2]{IFISC (UIB-CSIC), Instituto de Fisica Interdisciplinar y Sistemas Complejos - Palma de Mallorca, Spain}

\affil[*]{stefano.longhi@polimi.it}

\begin{abstract}
In waveguide quantum electrodynamics systems, atomic radiation emission is shaped by the photonic environment and collective atom interactions, offering promising applications in quantum technologies. In particular, atom-photon bound states, inhibiting complete spontaneous decay of the atom, can be realized through waveguide dispersion engineering or by utilizing giant atoms. While steady-state bound states are well understood, transient or virtual bound states remain less explored. Here we investigate transient atom-photon bound states, arising from initial atom-photon entanglement, and propose methods to slow down spontaneous atomic decay.
\end{abstract}

\setboolean{displaycopyright}{false} 

\begin{document}

\maketitle

{\em Introduction.}  It is widely acknowledged that the emission of radiation from atoms is not an intrinsic property of the atoms themselves, but rather is strongly influenced by the properties of the surrounding photonic environment and the collective interactions between atoms \cite{R1,R2,R3,R4,R5,R6,R7}. This result is particularly relevant in waveguide quantum electrodynamics (QED) systems, where one or more physical or artificial atoms are coupled to a one-dimensional (1D) waveguide \cite{R8,R9,R10,R11,R11b}. These systems offer exciting opportunities to explore strong light-matter interactions at the single-photon level, with promising applications in cutting-edge quantum technologies such as quantum communication, computation, and sensing \cite{R8,R11,R11b}.
The coupling of quantum emitters to 1D waveguides offers a highly controlled environment, enabling the investigation of novel phenomena such as quantum state transfer, photon spontaneous emission control, and entanglement generation.  When quantum emitters are coupled to waveguides with finite bandwidth or exhibit a colored interaction, atom-photon bound states can emerge \cite{R12,R13,R14,R15,R16,R17,R18,R19,R20,R21,R22,R23,R24,R24b,R24c,R25,R26,R27,R28,R28b,R29,R30,R31,R32,R33}. Bound states can exist both outside and within the continuum of propagating waveguide modes. These phenomena can be realized by carefully engineering waveguide dispersion, such as by using arrays of optical resonators \cite{
R16,R17,R18,R27}, by incorporating 1D topological photonic baths  \cite{R21,R23,R25}, or using artificial giant atoms \cite{R19,R24,R28,R29,R30,R33}. In such systems, the reduced dynamics of the atoms is usually described assuming that the photon field and the atoms are initially uncorrelated and in a disentangled state, with the photon field in the vacuum or in a thermal state. This assumption, however, is not always guaranteed in realistic scenarios and several theoretical approaches have been introduced  to characterize initially correlated dynamics of open quantum systems (see e.g. \cite{Referee1,Referee2,Referee3,Referee4}).  Initial system-environment correlations yield intriguing anomalous dynamical features, such as information backflow, strong non-Markovian behaviors and distance growth of quantum states.\\
In this letter, we predict the existence of transient atom-photon bound states in waveguide QED systems, arising from the initial preparation of the system in suitably entangled atom-photon states.
A transient atom-photon bound state corresponds to a bound state in which the emitter does not decay; however, it exists only for a finite time interval, after which the emitter undergoes irreversible decay into the continuum of propagating photonic modes. The existence of such virtual states could significantly influence the temporal evolution of light-matter interactions, making them relevant for controlling short-time dynamics, photon emission processes, and the transient behavior of waveguide QED systems.Additionally, we propose methods for controlling and slowing down atomic decay by carefully engineering the initial atom-photon state. As an illustrative example, we consider a two-level atom coupled to an array of coupled resonators, providing a framework to investigate transient atom-photon bound states and control spontaneous decay through initial atom-photon entanglement.
\\
{\em Model and decay dynamics.}  We consider the spontaneous emission of a point-like two-level quantum emitter which is electric-dipole coupled to the bosonic modes of a one-dimensional waveguide [Fig.1(a)].
 Indicating by $|e\rangle$ and $|g \rangle$ the excited and ground states of the emitter, with transition frequency $\omega_0$, and by $\omega(k)$ the dispersion relation of the waveguide modes with wave number $k$, in the rotating-wave approximation the Hamiltonian of the atom-photon field reads 
\begin{equation}
H=\omega_0 |e \rangle \langle e|+ \int dk \omega(k) a^{\dag}_ka_k+ \int dk \left\{  g(k) a_k |e \rangle \langle g|+{\rm H.c.} \right\}
\end{equation}
where  $a_k$ ($a^{\dag}_k$) are the photon annihilation (creation) operators of the waveguide modes, satisfying the usual bosonic commutation relations $[a_k,a_{k'}]=[a^{\dag}_k,a^{\dag}_{k'}]=0$ and $[a_k,a^{\dag}_{k'}]=\delta(k-k')$, and $g(k)$ is the atom-photon coupling constant. In the following analysis, we will assume a non-chiral waveguide, with the symmetry $\omega(-k)=\omega(k)$ and $g(-k)=g(k)$ for the dispersion curve and spectral coupling factor [Fig.1(b)].\\
Since $H$ commutes with $N=|e \rangle \langle e| + \int dk a^{\dag}_k a_k$, the total number of excitations $N$ is conserved. Assuming that the photon field is at zero temperature, to study the process of spontaneous emission we can restrict the analysis considering the subspace with excitations $N=1$ solely. In this case the atom-photon state is described by the wave function
\begin{equation}
|\psi(t) \rangle= \exp(-i \omega_0 t) \left\{ c_a(t) |e \rangle \otimes |0 \rangle + \int dk \varphi(k,t) |g \rangle \otimes a^{\dag}_k |0 \rangle \right\}
\end{equation}
where $|0 \rangle$ is the vacuum state of the photon field, $c_a(t)$ is the amplitude probability to find the atom in the excited state at time $t$, and $\varphi(k,t)$ is the amplitude probability that a photon has been emitted at time $t$ in the $k$-mode, with $|c_a(t)|^2+\int dk |\varphi(k,t)|^2=1$. The amplitude probabilities satisfy the coupled equations
\begin{eqnarray}
i \frac{dc_a}{dt} & = & \int dk g(k) \varphi(k,t) \\
i \frac{\partial \varphi}{\partial t} & = & \Omega(k) \varphi(k,t)+ g^*(k) c_a(t) 
\end{eqnarray}
where we have set $\Omega(k)=\omega(k)-\omega_0$.  The energy spectrum $E$ of the Hamiltonian $H$ in the single excitation sector $N=1$ is obtained from the spectral problem associated to Eqs.(3) and (4), which was solved in pioneering works by  Fano, Friedrichs and Lee \cite{R34,R35,R36}. Atom-photon bound states, either inside or outside the continuum, correspond to the point spectrum of $H$ and the conditions for their existence have been discussed in several previous works (see e.g. \cite{R13,R17,R33}). Here we assume a featureless continuum of modes, i.e. we do not consider mirror and/or finite-size effects of the waveguide,  and the weak atom-photon coupling regime, such that there are not atom-photon bound states and any initial excitation of the emitter irreversibly decays into the continuum of bosonic modes. 
The initial condition, $c_a(0)$ and $\varphi(k,0) \equiv \varphi_0(k)$, defines the atom-photon state at initial time $t=0$. Usually, one assumes the emitter in the excited state $|e \rangle$ and the photon field in the vacuum state, corresponding to the initial condition $c_a(0)=1$ and $\varphi_0(k)=0$. In this case the decay is described by a near exponential law with a decay rate $\gamma_R$ given by the Fermi golden rule, according to the Weisskopf-Wigner theory of spontaneous decay. Here we consider, conversely, an initial atom-photon entangled state, corresponding to $c_a(0), \varphi_0(k) \neq0$, and unveil how special initial states can give rise to a transient (or virtual) atom-photon bound state, or a near exponential spontaneous decay with a decay rate that can be far smaller than the golden rule value $\gamma_R$, i.e. to spontaneous emission slow down. For an initial atom-photon entangled state, the decay behavior of the emitter is described by the following {\em exact} integro-differential equation for the amplitude probability $c_a(t)$
\begin{equation}
\frac{dc_a}{dt}=F(t)- \int_0^t dt' \mathcal{G}(t-t') c_a(t')
\end{equation}
which is obtained from Eqs.(3) and (4) after elimination of $\varphi(k,t)$ from the dynamics. In the above equation, the memory function $\mathcal{G}(\tau)$ and forcing term $F(t)$ are given by
\begin{eqnarray}
\mathcal{G}(\tau) & = & \int dk |g(k)|^2 \exp[-i \Omega(k) \tau] \\
F(t) & = & -i \int dk g(k) \varphi_0(k) \exp[-i \Omega(k) t]. 
\end{eqnarray}
\begin{figure}[h]
 \centering
   \includegraphics[width=0.48\textwidth]{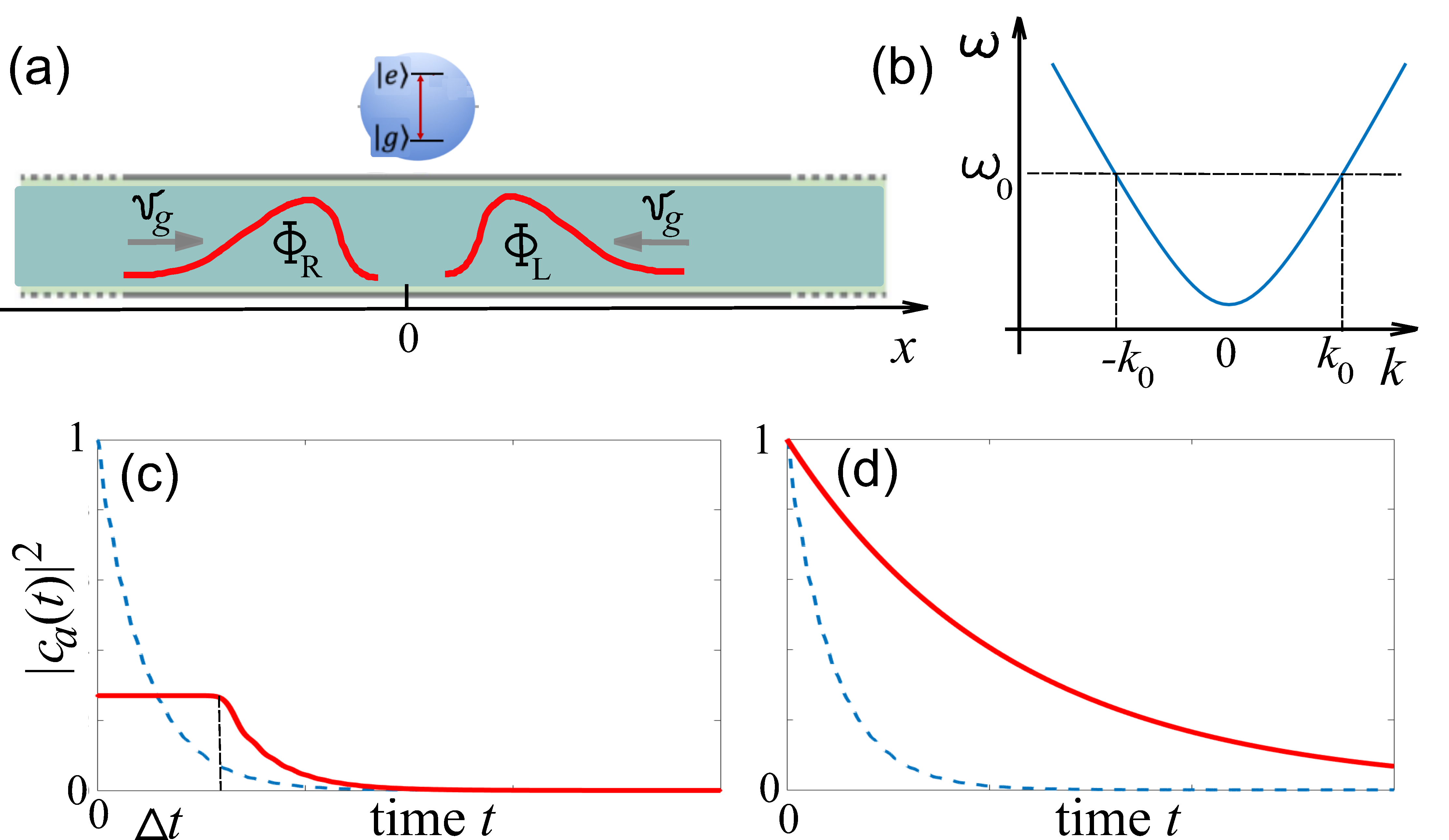}
   \caption{ \small (a) Schematic of an infinitely-extended waveguide with a two-level atom (resonance frequency $\omega_0$) placed at the position $x=0$. (b) Dispersion diagram $\omega=\omega(k)$ of the waveguide modes. Superposition of plane waves with wave numbers $k$ close to $k_0$ ($-k_0$) corresponds to a right-moving (left-moving) photon wave packet $\Phi_R(x)$ ($\Phi_L(x)$) with a group velocity $v_g$, as schematically shown by the red curves in panel (a). (c) Schematic of the temporal evolution of the probability $|c_a(t)|^2$ to find the atom in the excited state $|e \rangle$ corresponding to a virtual atom-photon bound state (bold red curve), and for the ordinary spontaneous emission decay (blue dashed curve) at the rate $\gamma_R$ given by the golden rule. For the virtual atom-photon bound state the probability remains stationary for a time interval $\Delta t$, after which spontaneous emission a the rate $\gamma_R$  occurs. (d) Same as (c), except that the initial atom-photon bound state leads to a near exponential decay at a rate $\epsilon$ much smaller than the golden rule value $\gamma_R$.}
 \end{figure}
From the stationary phase method it follows that $\mathcal{G}(\tau)$ decays to zero as $\tau \rightarrow  \infty$ with a characteristic time $\tau_m$ which defines the memory time of the non-Markovian dynamics. Under the weak coupling limit and assuming a short memory time, such that $c_a(t)$ varies slowly over the fast time scale $\tau_m$, we can introduce the markovian approximation replacing Eq.(5) with the following differential equation
\begin{equation}
\frac{dc_a}{dt} \simeq F(t)- \gamma c_a(t)
\end{equation}
where $\gamma=\int_0^ {\infty} d \tau \mathcal{G}( \tau) d \tau= \gamma_R+i \gamma_I$
 accounts for both the spontaneous emission decay rate $\gamma_R$ and Lamb frequency shift $\gamma_I$. Assuming a profile of the dispersion curve $\omega(k)$ of the waveguide modes as in Fig.1(b) and linearizing the dispersion relation around the wave numbers $k= \pm k_0$, where $\omega( \pm k_0)=\omega_0$ \cite{R8,Fan}, the decay rate and frequency shift read
 \begin{equation}
 \gamma_R=\frac{2 \pi |g_0|^2}{v_g} \; , \;\; \gamma_I= \mathcal{P} \int dk \frac{|g(k)|^2}{\omega_0-\omega(k)}.
 \end{equation}
 where $v_g=|(d \omega/ dk)_{\pm k_0}|$ is the group velocity of the waveguide modes and $g_0=g(\pm k_0)$ the coupling constant at the resonance frequency $\omega_0$.  Clearly, when the photon field is initially in the vacuum state, i.e. $c_a(0)=1$ and $\varphi_0(k)=0$, the forcing term $F(t)$ in Eq.(8) vanishes and one obtains the usual exponential decay law for $c_a(t)$ with the decay rate $\gamma_R$. However, an initial atom-photon entangled state $|\psi(0) \rangle= c_a(0) | e \rangle \otimes |0 \rangle + |g \rangle \otimes | \psi \rangle_{ph}$ with $| \psi \rangle_{ph}=\int \varphi_0(k) a^{\dag}_k |0 \rangle$,  can deeply modify the decay behavior owing to the non-vanishing forcing term $F(t)$ in Eq.(8). As shown below, suitable engineering of the initial atom-photon state can result in intriguing dynamical behaviors, such as the appearance of a transient atom-photon bound state, or spontaneous decay at a rate $\epsilon$ considerably smaller than $\gamma_R$. The main idea is to engineer the initial photon state $| \psi \rangle_{ph}$  such as to generate a target driving term $F(t)$ in Eq.(8), with $F(t)=0$ for $t<0$. As shown in the Supplemental document and schematically indicated in Fig.1(a), for the simple case of a waveguide sustaining plane-wave modes $\exp(ikx)$ with the quantum emitter placed at the waveguide position $x=0$, the initial photon state  $| \psi \rangle_{ph}$ yielding a target driving function $F(t)$ is not unique and corresponds rather generally to two counter-propagating wave packets, a right-propagating wave packet $\Phi_R(x)$ and a left-propagating wave packet $\Phi_L(x)$. In particular, one can assume a single wave packet, the right-propagating wave packet. In this case, as shown in the Supplemental document the initial photon wave packet is fully localized on the left side of the emitter and has a spatial envelope profile given by $\Phi_R(x) \propto F(- x/v_g)$.\\
 {\em Transient atom-photon bound state.}  A virtual atom-photon bound state corresponds to a  state where transient suppression of atom decay into the continuum is observed. Quantitatively, for a virtual state $|c_a(t)|^2$ is constant for a given time interval $\Delta t$, namely $|c_a(t)|^2=|c_a(0)|^2$ for $0 \leq t < \Delta t$, after which an irreversible decay of $c_a(t)$ to zero is observed [Fig.1(c)]. Note however that, while the dynamics of the atom is frozen in the time interval $0<t<\Delta t$, the photon wave function $\varphi(k,t)$ does evolve in this time interval. 
  A virtual atom-photon state is obtained by engineering the initial atom-photon state such that $F(t)$ is a stepwise function, with $F(t)=\gamma c_a(0)$ for $0<t<\Delta t$ and $F(t)=0$ for $t> \Delta t$. The explicit expression of $\varphi_0(k)$, i.e. initial atom-photon entangled state, is given in the Supplemental document.  To observe the virtual atom-photon bound state for a time interval $\Delta t$, one can consider a single initial photon wave packet. As shown in the Supplemental document, the initial amplitude $c_a(0)$ should satisfy the condition $ |c_a(0)|^2= 1/(1+ \gamma_R \Delta t)$, indicating that $|c_a(0)|^2 \ll 1$ for a trapping time $\Delta t$ much larger than the spontaneous lifetime $1/ \gamma_R$. This result is consistent with the fact that there are not stationary atom-photon bound states, and thus a long-lived virtual atom-photon bound state should correspond to an initial vanishing probability to find the atom in the excited state. \\
{\em Spontaneous emission slow down.}  The second intriguing dynamical behavior corresponds to a nearly exponential decay law, like in the ordinary Weisskopf-Wigner theory, but with a decay rate $\epsilon$ much smaller than the value $\gamma_R$ given by the golden rule [Fig.1(d)]. This regime is obtained by engineering the initial atom-photon state such that $F(t)=\gamma c_a(0) \exp(-\epsilon t)$. In fact, when the driving term is chosen in this way, the solution to Eq.(8) reads
\begin{eqnarray}
c_a(t) & = & c_a(0) \left\{ \frac{\epsilon}{\epsilon-\gamma} \exp(-\gamma t)+\frac{\gamma}{\gamma-\epsilon} \exp(- \epsilon t)  \right\} \nonumber \\
&  \simeq & c_a(0) \exp(- \epsilon t)
\end{eqnarray}
indicating a near exponential decay at the rate $\epsilon \ll \gamma_R$. The explicit form of the spectral amplitude $\varphi_0(k)$ generating the driving term $F(t)=\gamma c_a(0) \exp(-\epsilon t)$ is given in the Supplemental document. The initial probability $|c_a(0)|^2$ to find the emitter in the excited state is related to $\epsilon$ by the relation $ |c_a(0)|^2=1/(1+ \gamma_R/ 2 \epsilon)$,
indicating that $c_a(0) \rightarrow 0$ as $\epsilon / \gamma_R \rightarrow 0$. \\ 
\begin{figure}
 \centering
   \includegraphics[width=0.48\textwidth]{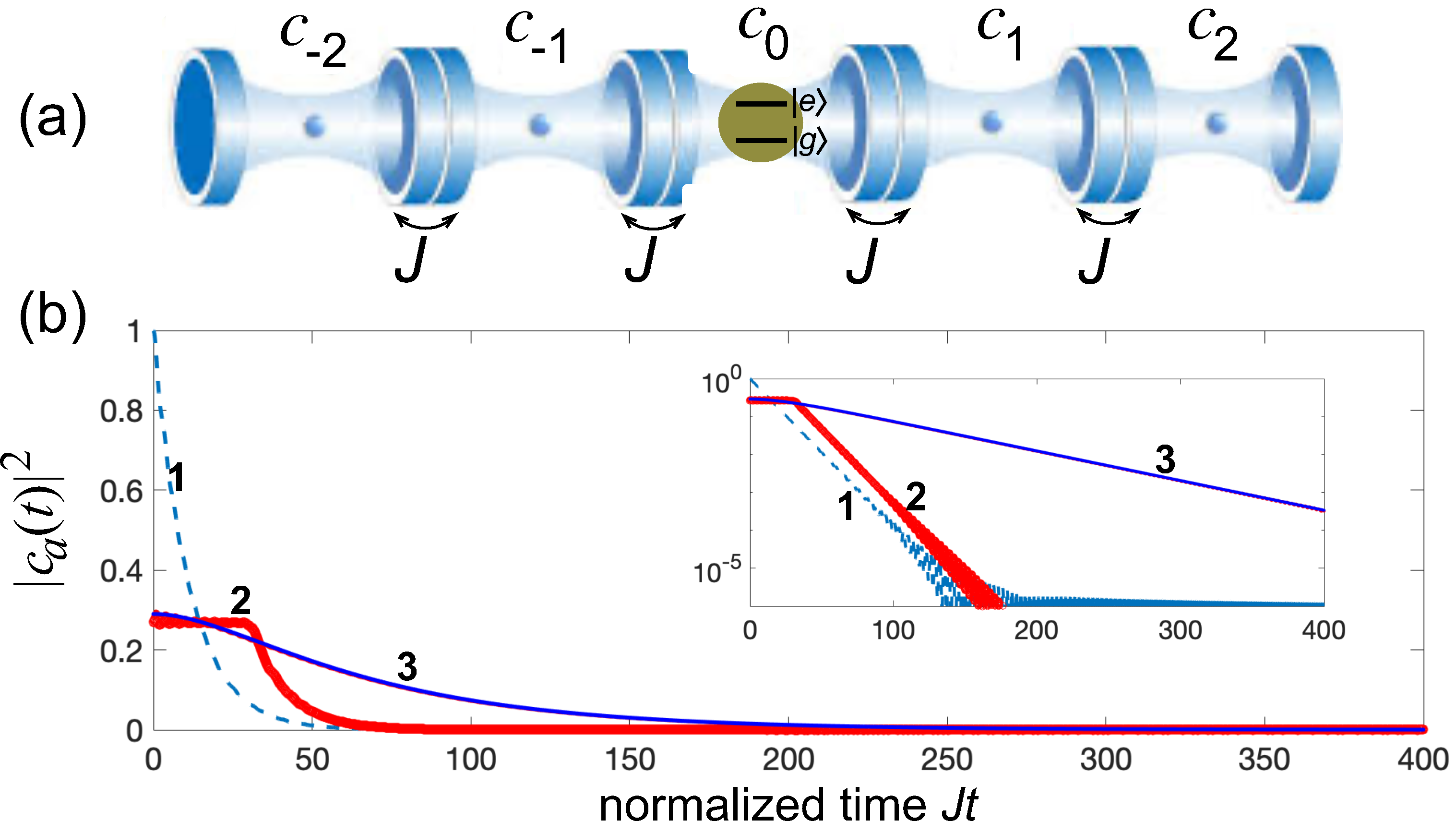}
   \caption{ \small (a) Schematic of an array of coupled optical cavities (coupling constant $J$) with a two-level atom placed inside one of the resonators. The atom-photon coupling rate is $G_0$. (b) Numerically-computed temporal evolution of the atom survival probability $|c_a(t)|^2$ for different initial atom-photon states $| \psi(0) \rangle= c_a(0) |e \rangle \otimes |0 \rangle+|g \rangle \otimes \sum_l  Q_l c^{\dag}_l |0 \rangle$ for $G_0/J=0.3$. The waveguide is formed by a large enough number of cavities ($2N+1=1601$) to avoid edge effects up to the observation time.  Curve 1 is the ordinary near-exponential spontaneous emission decay for the initial separable state $|\psi(0) \rangle =|e \rangle \otimes |0 \rangle$. The curve is very well fitted by an exponential profile with the golden decay rate $\gamma_R=G_0^2/(2J)$. Curve 2 is the decay dynamics corresponding to the virtual atom-photon state for $ \Delta t=30/J$ [the initial entangled state is given by Eqs.(12) and (14)]. Curve 3 corresponds to spontaneous emission slow down, obtained by assuming the initial entangled state defined by Eqs. (12) and (15) with $\epsilon=\gamma_R/5$. The curve is fitted with excellent agreement with the theoretical prediction [Eq.(10)].
   The inset depicts the decay behavior of the three curves on a log vertical scale. }
 \end{figure}
{\em Illustrative example.} To illustrate the above dynamical regimes, let us assume that the photonic waveguide is an array of optical resonators \cite{R16,R18,R24,R27,R28b,R31}, as schematically shown in Fig.2(a). In Wannier basis, the full Hamiltonian of the system reads 
\begin{eqnarray}
H & = & \omega_0 |e \rangle \langle e|+ \sum_l  \left\{\omega_c c^{\dag}_l c_l-J(c^{\dag}_{l+1}c_l+{\rm H.c.}) \right\} \nonumber \\
&+ & G_0 \left( c_0^{\dag} |g \rangle \langle e| +{\rm H.c.} \right)
\end{eqnarray}
where $\omega_c \simeq \omega_0$ is the resonance frequency of the electromagnetic mode  in each single resonator, $J$ is the photon hopping rate between adjacent resonators, $c^{\dag}_l$ ($c_l$) is the creation (destruction) operator of the photon field in the $l$-th resonator of the array (Wannier basis), and $G_0$ is the electric-dipole coupling strength of the two-level atom placed in the $l=0$ cavity. The Hamiltonian (11) can be cast in the canonical form Eq.(1), with $\omega(k)=\omega_c-2J \cos(k)$ and $g(k)=G_0/ \sqrt{2 \pi}$, after introduction of the creation/destruction operators of Bloch modes of the photonic lattice, i.e. after letting $a_k=(1/ \sqrt{2 \pi}) \sum_l c_l \exp(-ikl)$, where $-\pi \leq k < \pi$ is the Bloch wave number. Assuming the resonance condition $\omega_c=\omega_0$, corresponding to $k_0= \pi/2$, the group velocity of the two counter-propagating waveguide modes at wave numbers $\pm k_0$ is $v_g=2J$. The weak coupling regime corresponds to $G_0 \ll J$. From Eq.(9), the golden rule decay rate and Lamb frequency shift can be readily computed, yielding  $\gamma_R= G_0^2/(2J)$ and $\gamma_I=0$. The initial entangled atom-photon state can be written in Wannier basis as 
\begin{equation}
|\psi(0) \rangle=c_a(0) |e \rangle \otimes |0 \rangle+ \sum_l Q_l |g \rangle \otimes \hat{c}^{\dag}_l |0 \rangle 
\end{equation}
where the complex amplitudes $Q_l$ are given in terms of  the spectral function $\varphi_0(k)$ via the relations
$Q_l= 1/{\sqrt{2 \pi}} \int dk \varphi_0(k) \exp(ikl)$. The driving function $F(t)$ can be expressed as a von-Neumann series with coefficients $Q_l$, namely (technical details are given in the Supplemental document)
\begin{equation}
F(t)=G_0 \sum_{l=-\infty}^{\infty} Q_l J_l(2 J t) \exp[i k_0 (l-1)],
\end{equation}
where $J_l(x)$ is the Bessel function of first kind and order $l$.  Owing to the oscillatory behavior of the Bessel functions, $F(t)$ displays rapid oscillations on the fast time scale $\sim  2 \pi /J$. However, since in the weak coupling regime $c_a(t)$ cannot follow such fast oscillations, one can replace the exact form of $F(t)$, given by Eq.(13), with a local time-average $\overline {F(t)}$, over a time scale $2 \pi/J$.  
As discussed in the Supplemental document, the virtual atom-photon bound state is obtained  by assuming, for example
\begin{equation}
Q_l=
\left\{
\begin{array}{cc}
i c_a(0) (G_0 / 2J) \exp(-ik_0  l  ) & |l| \leq L \\
0 & {\rm otherwise }
\end{array}
\right.
\end{equation}
 with $\Delta t= L/v_g=L/(2J)$ and  $|c_a(0)|^2 \simeq (1+G_0^2 L/ 2J^2)^{-1}$ Likewise, spontaneous emission slow down, at a rate $\epsilon$ much smaller than the golden rule value $\gamma_R$, can be observed assuming
\begin{equation}
Q_l=   
\left\{
\begin{array}{cc}
i c_a(0) ( G_0/2J) \exp(- \epsilon l /v_g-ik_0 l) & l \ge 0 \\
0 & l<0
\end{array}
\right.
\end{equation}
with the normalization condition $|c_a(0)|^2 \simeq 1/(1+\gamma_R / 2 \epsilon)$. We checked the existence of both  virtual atom-photon bound states and spontaneous emission slow down, predicted by the above theoretical analysis, by exact  computation of the decay profile $|c_a(t)|^2$ numerically solving the Schr\"odinger equation $i \partial_t | \psi(t) \rangle = H | \psi(t) \rangle$ in the one excitation sector, i.e. beyond the markovian approximation used to derive Eq.(8). Typical numerical results are depicted in Fig.2(b) for parameter values $\omega_0=\omega_c$ and $G_0/J=0.3$. The waveguide comprises $(2N+1)$ cavities, and the emitter is placed in the central cavity. Curve 1 in the figure shows the decay behavior of $|c_a(t)|^2$ for the conventional initial condition $c_a(0)=1$ and $\varphi_0(k)=0$ ($Q_l=0$), leading to a near exponential decay with a rate in excellent agreement with the golden rule value $\gamma_R=G_0^2/(2J)=0.045J$. Curve 2 corresponds to the numerically-computed decay curve $|c_a(t)|^2$ as obtained assuming Eqs.(12,14) as an initial condition with $L=60$, corresponding to a virtual atom-photon bound state surviving for the time interval $\Delta t= L/(2J)=30/J$. Finally, curve 3 depicts the numerically-computed decay curve $|c_a(t)|^2$ as obtained assuming Eqs.(12,15) as an initial condition, clearly indicating spontaneous emission slow down. A possible route for a feasible preparation of the system in an entagled atom-photon bound state, suited for the observation of virtual bound states, is presented in Sec.4. of the Supplemental document.\\
{\em Conclusion.}
In this work, we have explored the existence of transient atom-photon bound states in waveguide QED systems, focusing on the role of initial atom-photon entanglement in shaping the temporal dynamics of radiation emission. We have shown that these virtual bound states can affect short-time light-matter interactions, leading to a transient frozen dynamics, and proposed strategies to control and slow down atomic decay through careful engineering of the atom-photon state. Our findings could be of significant interest for quantum applications that require precise control over photon dynamics and atom-photon interactions.\\
\\ 
\noindent
{\bf Disclosures}. The author declares no conflicts of interest.\\
{\bf Data availability}. No data were generated or analyzed in the presented research.\\
{\bf Funding}. Agencia Estatal de Investigacion (MDM-2017-0711).\\
{\bf Supplemental document}. See Supplement 1 for supporting content.

\newpage


 {\bf References with full titles}\\
 \\
 \noindent
 1. S. Haroche and J.M. Raimond, {\em Exploring the Quantum: Atoms, Cavities, and Photons} (Oxford University Press, 2006).\\
 2.  R.H. Dicke, Coherence in spontaneous radiation processes, Phys. Rev. {\bf 93}, 99 (1954).\\
 3. M. Gross and S. Haroche, Superradiance: An essay on the theory of collective spontaneous emission, Phys. Rep. {\bf 93}, 301 (1982).\\
 4. P. Goy, J. M. Raimond, M. Gross, and S. Haroche, Observation of cavity-enhanced single-atom spontaneous emission,  Phys. Rev. Lett. {\bf 50}, 1903 (1983).\\
 5. R.J. Cook and P.W. Milonni, Quantum theory of an atom near partially refiecting walls, Phys. Rev. A {\bf 35}, 5081
(1987).\\
 6. E. Yablonovitch, Inhibited spontaneous emission in solid-state physics and electronics, Phys. Rev. Lett. {\bf 58}, 2059 (1987).\\
 7. S. Noda, M. Fujita, and T.  Asano, Spontaneous-emission control by photonic crystals and nanocavities, Nat. Photon. {\bf 1}, 449 (2007).\\
 8. D. Roy, C.M. Wilson, and O. Firstenberg, Colloquium: Strongly interacting photons in one-dimensional continuum,
Rev. Mod. Phys. {\bf 89}, 021001 (2017).\\
 9.  X. Gu, A. F. Kockum, A. Miranowicz, Y.-x. Liu, and F. Nori, Microwave photonics with superconducting quantum circuits, Phys.
Rep. {\bf 718-719}, 1 (2017).\\ 
 11. B. Kannan, M. J. Ruckriegel, D. L. Campbell, A.F. Kockum, J. Braum\"uller, D.K. Kim, M. Kjaergaard, P. Krantz, A. Melville, B.M. Niedzielski, A. Veps\"al\"ainen, R. Winik, J.L. Yoder, F. Nori, T.P. Orlando, S. Gustavsson, and W.D. Oliver, Waveguide quantum electrodynamics with superconducting artificial giant atoms, Nature {\bf 583} 775 (2020).\\
11. A.S. Sheremet, M.I. Petrov, I.V. Iorsh, A.V. Poshakinskiy, and A.N. Poddubny,
 Waveguide quantum electrodynamics: Collective radiance and photon-photon correlations,
Rev. Mod. Phys. {\bf 95}, 015002 (2023).\\
12. A. Gonz\'alez-Tudela, A. Reiserer, J.J. Garcia-Ripoll, and F.J Garcia-Vidal, Light-matter interactions in quantum nanophotonic devices, 
Nature Rev. Phys. {\bf 6},166 (2024).\\
13. S. John and J. Wang, Quantum electrodynamics near a photonic band gap: Photon bound states and dressed atoms, Phys. Rev. Lett. {\bf 64}, 2418 (1990).\\
14. S. Longhi, Bound states in the continuum in a single-level Fano-Anderson model, Eur. Phys. J. B {\bf 57}, 45 (2007).\\
15. T. Tufarelli, F. Ciccarello, and M. S. Kim, Dynamics of spontaneous emission in a single-end photonic waveguide, Phys. Rev. A {\bf 87}, 013820 (2013).\\
16. T. Tufarelli, M.S. Kim, and F. Ciccarello, Non-Markovianity of a quantum emitter in front of a mirror, Phys. Rev. A {\bf 90}, 012113 (2014).\\
17. F. Lombardo, F. Ciccarello, and G.M. Palma, Photon localization versus population trapping in a coupled-cavity array, Phys. Rev. A {\bf 89}, 053826 (2014).\\
18. T. Shi, Y.-H. Wu, A. Gonz\'alez-Tudela, and J. I. Cirac, Bound states in boson impurity models, Phys. Rev. X {\bf 6}, 021027 (2016).\\
19. G. Calaj\`o, F. Ciccarello, D. Chang, and P. Rabl, Atom-field dressed states in slow-light waveguide QED, Phys. Rev. A
{\bf 93}, 033833 (2016).\\
20. A.F. Kockum, G. Johansson, and F. Nori, 
Decoherence-Free Interaction between Giant Atoms in
Waveguide Quantum Electrodynamics,
Phys. Rev. Lett. {\bf 120}, 140404 (2018).\\
21. P. Facchi, D. Lonigro, S. Pascazio, F.V. Pepe, and D. Pomarico,
Bound states in the continuum for an array of quantum emitters,
Phys. Rev. A {\bf 100}, 023834 (2019).\\
22. M. Bello, G. Platero, J.I. Cirac, A. Gonz\'alez-Tudela, Unconventional quantum optics in topological waveguide QED, Sci. Adv. {\bf 5}, eaaw0297 (2019).\\
23. G. Calaj\'o, Y.-L. L. Fang, H.U. Baranger, and F. Ciccarello,  Exciting a Bound State in the Continuum through Multiphoton Scattering Plus Delayed Quantum Feedback,
Phys. Rev. Lett. {\bf 122}, 073601 (2019).\\
24. S. Mahmoodian, G. Calaj\'o, D.E. Chang, K. Hammerer, and A.S. S{\o}rensen, 
Dynamics of Many-Body Photon Bound States in Chiral Waveguide QED, Phys. Rev. X {\bf 10}, 031011 (2020).\\
25. S. Longhi, Photonic simulation of giant atom decay, Opt. Lett. {\bf 45}, 3017 (2020).\\
26. J. Rom\'an-Roche, E. S\'anchez-Burillo, and D. Zueco, Bound states in ultrastrong waveguide QED,
Phys. Rev. A {\bf 102}, 023702 (2020).\\
27. K. Sinha, P. Meystre, E.A. Goldschmidt, F.K. Fatemi, S.L. Rolston, and  P. Solano, Non-Markovian Collective Emission from Macroscopically Separated Emitters,
Phys. Rev. Lett. {\bf 124}, 043603 (2020).\\
28. E. Kim, X. Zhang, V.S. Ferreira, J. Banker, J.K. Iverson, A. Sipahigil, M. Bello, A. Gonz\'alez-Tudela,  M. Mirhosseini, and O. Painter, Quantum electrodynamics in a topological waveguide, Phys. Rev. X {\bf 11}, 011015 (2021).\\
29. S. Lorenzo, S. Longhi, A. Cabot, R. Zambrini, and G.L. Giorgi, Intermittent decoherence blockade in a chiral ring environment, Sci. Rep. {\bf 11}, 12834 (2021).\\
30. M. Scigliuzzo, G. Calaj\`o, F. Ciccarello, D. Perez Lozano, A. Bengtsson, P. Scarlino, A. Wallraff, D. Chang, P. Delsing, and S. Gasparinetti, Controlling atom-photon bound states in an array of Josephson Junction resonators, Phys. Rev. X {\bf 12}, 031036 (2022).\\
31. A, Soro and A.F. Kockum, Chiral quantum optics with giant atoms, Phys. Rev. A {\bf 105}, 023712 (2022).\\
32. H. Xiao, L. Wang, Z.-H. Li, X. Chen, and L. Yuan, Bound state in a giant atom-modulated resonators system, 
npj Quant. Inf. {\bf 8}, 80 (2022).\\ 
33. N. Tomm, S. Mahmoodian, N. O. Antoniadis, R. Schott, S.R. Valentin, A.D. Wieck, A. Ludwig, A. Javadi, and R.J. Warburton, 
Photon bound state dynamics from a single artificial atom, Nature Phys. {\bf 19}, 857 (2023).\\
34.W.Z. Jia and M.T. Yu,  Atom-photon dressed states in a waveguide-QED system with multiple giant
atoms, Opt. Express {\bf 32}, 9495 (2024).\\
35. Z. Lu, J. Li, J. Lu, and L. Zhou, Controlling atom-photon bound states in a coupled resonator array with a two-level quantum emitter, Opt. Lett. {\bf 49}, 806 (2024).\\
36. W. Alvarez-Giron, P. Solano, K. Sinha, and P. Barberis-Blostein, 
Delay-induced spontaneous dark-state generation from two distant excited atoms,
Phys. Rev. Research  {\bf 6}, 023213 (2024).\\
37. L. Leonforte, X. Sun, D. Valenti, B. Spagnolo, F. Illuminati, A. Carollo, and F. Ciccarello, 
Quantum optics with giant atoms in a structured photonic bath, Quantum Sci. Technol. {\bf 10}, 015057 (2025).\\
38. A. Smirne, H.-P. Breuer, J. Piilo,, and B. Vacchini, Initial correlations in open-systems dynamics: The Jaynes-Cummings model,
Phys. Rev. A {\bf 82}, 062114 (2010).\\
 39. A.G. Dijkstra and Y. Tanimura, Non-Markovian Entanglement Dynamics in the Presence of System-Bath Coherence, Phys. Rev. Lett. {\bf 104}, 250401 (2010).\\
40. H.-T. Tan and W.-M. Zhang,  Non-Markovian dynamics of an open quantum system with initial system-reservoir
correlations: A nanocavity coupled to a coupled-resonator optical waveguide,  Phys. Rev. A {\bf 83}, 032102 (2011).\\
41. F. Settimo, K. Luoma, D. Chruscinski, A. Smirne, B. Vacchini, and J. Piilo,
Dynamics of Open Quantum Systems with Initial System-Environment Correlations via Stochastic Unravelings, arXiv:2502.12818 (2025).\\
42. U. Fano, Effects of Configuration Interaction on Intensities and Phase Shifts, Phys. Rev. {\bf 124}, 1866 (1961).\\
43. K.O. Friedrichs, On the perturbation of continuous spectra, Commun. Pure Appl. Math. {\bf 1}, 361 (1948).\\
44. T.D. Lee, Some Special Examples in Renormalizable Field Theory, Phys. Rev. {\bf 95}, 1329 (1954).\\
45. J.-T. Shen and S. Fan,
Theory of single-photon transport in a single-mode waveguide.
I. Coupling to a cavity containing a two-level atom, Phys. Rev. A {\bf 79}, 023837 (2009).

 \end{document}